\documentclass[showpacs,preprintnumbers,amsmath,amssymb,twocolumn]{revtex4}

\usepackage{graphicx}% Include figure files
\usepackage{dcolumn}% Align table columns on decimal point
\usepackage{bm}% bold math
\usepackage[usenames]{color}

\begin{document}

%\preprint{draft version}

\title{Efficient generation of random multipartite entangled states using time optimal unitary operations}
\author{A. Borras\footnote{toni.borras@uib.es}, A.P. Majtey\footnote{ana.majtey@uib.es} and
M. Casas\footnote{montse.casas@uib.es} } \affiliation{Departament
de F\'{\i}sica and IFISC-CSIC, Universitat de les Illes Balears,
07122 Palma de Mallorca, Spain }

\date{\today}
\begin{abstract}
We review the generation of random pure states using a protocol of
repeated two qubit gates. We study the dependence of the
convergence to states with Haar multipartite entanglement
distribution. We investigate the optimal generation of such states
in terms of the physical (real) time needed to apply the protocol,
instead of the gate complexity point of view used in other works.
This physical time can be obtained, for a given Hamiltonian,
within the theoretical framework offered by the quantum
brachistochrone formalism. Using an anisotropic Heisenberg
Hamiltonian as an example, we find that different optimal quantum
gates arise according to the optimality point of view used in each
case. We also study how the convergence to random entangled states
depends on different entanglement measures.
\end{abstract}


\maketitle

\section{Introduction}
One of the most fundamental concepts in the quantum description of
Nature is that of entanglement \cite{BZ06,PV07}. Entanglement
constitutes a physical resource that lies at the heart of
important information processes \cite{NC00,LPS98,BEZ00} such as
quantum teleportation, superdense coding, and quantum computation.
It has recently been proved that the generic entanglement, defined
as the entanglement of random states, can be produced in a
polynomial time using random two-qubit gates \cite{DOP07}.

Random states and random unitary operators are the quantum analogs
of random numbers and are two basic concepts in quantum
information and quantum communication tasks. Random unitary
operators are involved in the superdense coding of arbitrary
states \cite{HHL04} while the classical capacity of a noisy
quantum channel is saturated by random quantum states \cite{L97},
just to mention two very significant applications. Both of them
are well defined by the Haar measure which remains invariant under
unitary transformation. Since the generation of random states is
exponentially hard we study the production of states with the same
distribution of entanglement, a task that can be performed with
fewer physical resources \cite{ODP07}.

Although a large amount of different multipartite entanglement
measures has recently been proposed, a considerable amount of
research has particularly been devoted to the study of multiqubit
entanglement measures defined as the sum of bipartite entanglement
measures over all (or an appropriate family of) the possible
bi-partitions of the full system
\cite{WH05,CMB04,AM06,CHDB05,BPBZCP07}. There exist two popular
entanglement measures for multiqubit pure states, one based on the
von Neumann entropy of marginal density matrices and the other one
based upon the linear entropy of those matrices. It has recently
been shown that the von Neumann entropy based measure is able to
grasp more features of highly entangled states than the linear
entropy based measure \cite{BCPP08}. It is then expected that when
the efficient generation of random states is studied in the light
of these entanglement measures, some differences will arise
because, as a consequence of the ``concentration of measure"
phenomenon, these states are almost maximally entangled
\cite{HLW06}.

Emerson {\it et al}. \cite{EWSLC03} introduced a protocol for
generating pseudo-random unitary operators. A circuit of repeated
two-qubit gates was considered acting on a separable state of $N$
qubits. The unitary operation is given by the combination of two
independent single qubit rotations, chosen according to the
invariant Haar measure at each time step, and a fixed two-qubit
gate. The convergence to the Haar measure was shown to be
polynomial with the number of qubits. Some experimental evidences
have reinforced these results. A Markovian description of certain
two-qubit gates has also been used to analytically prove that the
convergence is reached in a polynomial time with the size of the
system \cite{ODP07}. Recently, a numerical effort has allowed to
identify the optimal two-qubit gate improving this analytical
bound \cite{Z07}.

In these previous works convergence was studied in terms of gate
complexity, i.e. the number of gates needed to reach the Haar
distribution. In contraposition to the gate complexity, a new
complexity concept for quantum algorithms has been proposed: the
time complexity \cite{SHSKG05,WJB02}, understood as the
physical time needed to perform such algorithm. The minimization
of this time is as important, from the experimental point of view,
as the gate complexity. A considerable amount of work has recently
been devoted to the time-optimal quantum computation problem, with
emphasis in the quantum Brachistochrone formalism
\cite{CHKO06,BH06,BZPCP08,CHKO07}. Making use of the analogy with
the brachistochrone problem from classic mechanics, Carlini et. al
\cite{CHKO06} introduced a variational approach to obtain the
optimal Hamiltonian and the optimal quantum  evolution between
initial and final given states. A geometric approach to solve this
problem, based on the symmetry properties of the quantum states
space, was addressed in \cite{BH06}. The role of the entanglement
within the quantum brachistochrone formalism was studied in
\cite{BZPCP08}.

A more general result has recently been formulated in terms of the
variational principle to find the time-optimal duration of a
unitary transformation \cite{CHKO07}. This formulation is
independent of the input state and because of that more general
than the previously described one. The time-optimal way to obtain
a two-qubit universal quantum gate was previously discussed using
the Cartan decomposition scheme for unitary transformation and
under the constraint that one-qubit gates can be performed
arbitrarily fast \cite{KG01, ZVSW03}. To study the brachistochrone
problem in relation with unitary operations, the quantum states
space is replaced with the space of unitary operators. This
formalism allows to consider the constraint imposed by the finite
amount of energy available in a physical experiment, as well as
any other constraints imposed by experimental requirements or
theoretical conditions. Then, the problem of finding the optimal
parameters for the Hamiltonian, is reduced to the resolution of a
set of ordinary differential equations \cite{CHKO07}.

The aim of this contribution is to study the possible differences
between both formalisms i.e. the gate and time complexities. We
also study the dependence of the convergence rate with different
entanglement measures. The paper is organized as follows.  In Sec.
II we review the protocol to generate random pure states and
discuss the dependence of the convergence time, in terms of the
number of gates to be applied, with the entanglement measure. In
Sec. III we study the physical time for the convergence. Finally,
Sec. IV is devoted to summarize and discuss our results.

\section{Generation of random bipartite entanglement}

The efficient quantum circuit generating random quantum states of
N qubits is based on the iterative application of a two-qubit
quantum gate $U_{ij}$ acting on qubits $i$ and $j$, arbitrarily
drawn from the N-qubit system, at each time step. The quantum gate
$U_{ij}$ is composed by the product of two single qubit rotation
gates $V_i$ and $V_j$ uniformly drawn from the Haar measure on U(2), and a fixed two qubit gate $W_{ij}$

\begin{equation}\label{entgate}
U_{ij} \, = \, V_i \, V_j \, W_{i,j},
\end{equation}

\noindent where $W_{ij}$ can be decomposed as

\begin{equation}\label{2qubit-decomp}
W_{ij} \, = \, (v_1 \otimes v_2) \, \exp \, [-\imath \sum_{k=x,y,z} \lambda_k \, \sigma_k
\, \otimes \, \sigma_k] \, (u_1 \otimes u_2),
\end{equation}

\noindent with the fixed rotations $v_{1,2}$ and $u_{1,2}$ acting only on
one of the two qubits, and $\sigma_k$ are the Pauli matrices
\cite{KBG01,KC01}. In our random entangling protocol the role of the fixed local rotations $u_{1,2}$ and $v_{1,2}$ should not be confused with the random local rotations $V_{i,j}$ , which change at each step of the protocol. As we are just interested in the entanglement generation properties of the two-qubit gates we only need to consider the action of the non local part of the decomposition (\ref{2qubit-decomp}), because its entangling power is the same when averaged over a large number of realizations \cite{KC01,BCPP05}. The symmetries
of such non-local action enables us to consider just a reduced
range for the values of its parameters ($\lambda_k \in
[0,\pi/4], \, k=x,y,z$). The qubits $i$ and $j$ upon which the gate $U_{ij}$
is applied can be chosen in several different ways, and each of
them corresponds to different geometries of the system: local and
non local \cite{MSB07}. In the non local case qubits $i$ and $j$ are chosen
randomly, the gate can act on two arbitrarily separated qubits. In
the local case the gate can only act on two neighboring qubits. In
this scheme we study both, periodic and open boundary conditions.

The results are qualitatively the same in the local and non local
case, so we only show those of the non local couplings as
representative of the typical behavior. The main difference is
that a larger amount of two-qubit entangling gates are needed to
converge to the entanglement of random states for the local
geometry.

%We combine a two-qubit gate with random single qubit rotation as
%in \cite{MSB07} and in the following we consider that the time to
%perform this latter operation is arbitrarily short. The optimal
%gate in terms of gate complexity according to each geometry was
%presented in \cite{Z07}.

Our goal is to reproduce the entanglement of typical random
states, to such an end we averaged a large enough number of
realizations (typically $10^3$) in order to have small statistical
fluctuations.

The genuine multipartite entanglement $E$ of a $N$-qubit state can
be expressed as

\begin{eqnarray}
E &=& \frac{1}{[N/2]} \sum_{m=1}^{[N/2]} E^{(m)}, \\
E^{(m)} &=& \frac{1}{N_{bipart}^m} \sum_{i=1}^{N_{bipart}^m} E(i).
\label{Entsub}
\end{eqnarray}
Here, $E(i)$ stands for the entanglement associated with one,
single bi-partition of the $N$-qubits system. The quantity
$E^{(m)}$ gives the average entanglement between subsets of $m$
qubits and the remaining $N-m$ qubits constituting the system. The
average is performed over the $N_{bipart}^{(m)}$ nonequivalent
ways to do such bi-partitions, which are given by

\begin{eqnarray}
N_{bipart}^{m} &=& \binom{N}{m}\textrm{     if }m \neq N/2,\\
N_{bipart}^{N/2} &=& \frac{1}{2} \binom{N}{N/2}\textrm{     if }m
= N/2.
\end{eqnarray}

\noindent Different $E^{(m)}$ represent different entanglement
properties of the state, this is why all these entanglement
measures must be computed to capture all the entanglement
properties of the state. The global multiqubit entanglement is
given by the average of the $[N/2]$ different $E^{(m)}$ for any
state $|\Psi\rangle$.

\begin{figure}
\begin{center}
\vspace{0.5cm}
\includegraphics[scale=0.35,angle=-90]{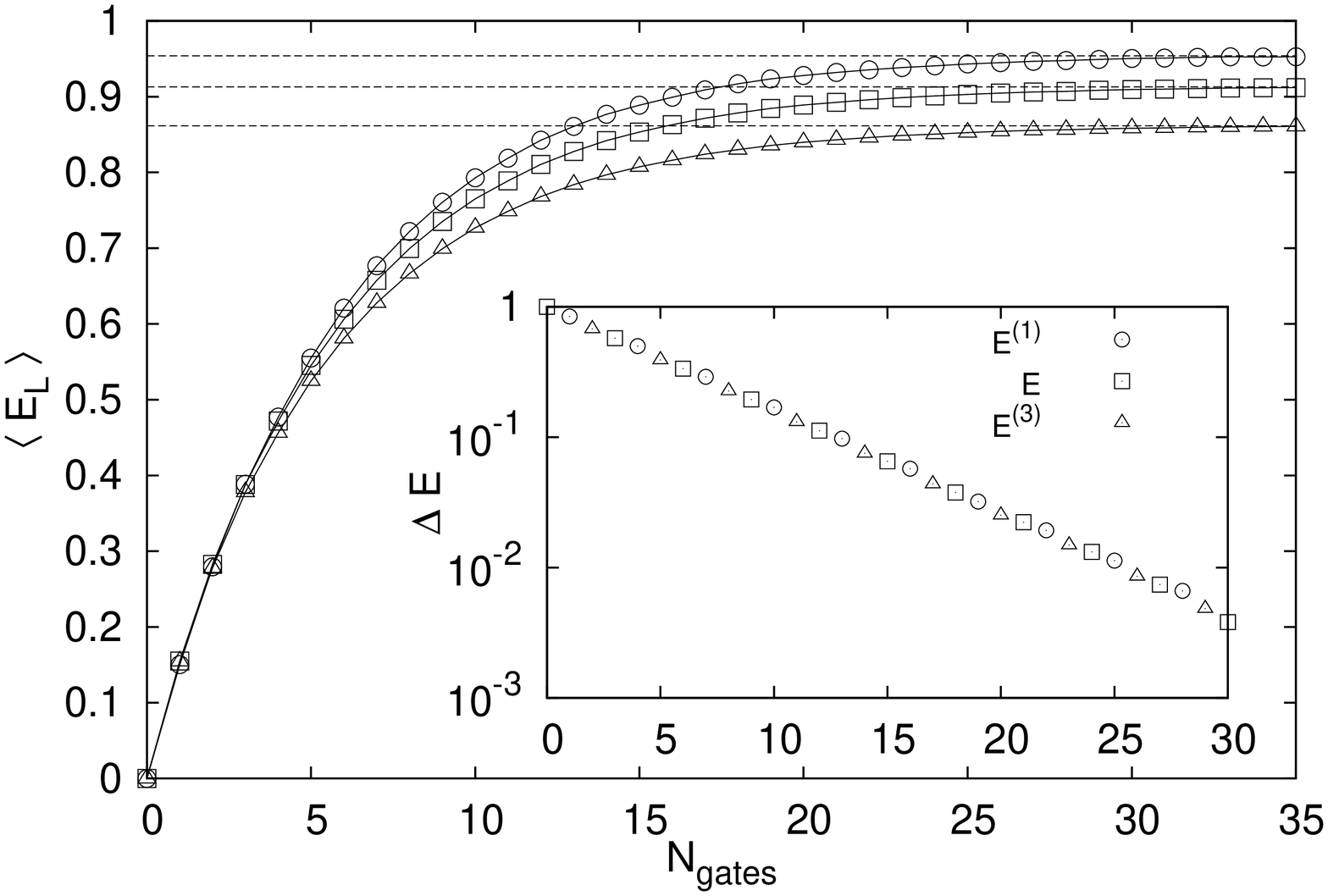}
\vspace {0.5cm}
\includegraphics[scale=0.35,angle=-90]{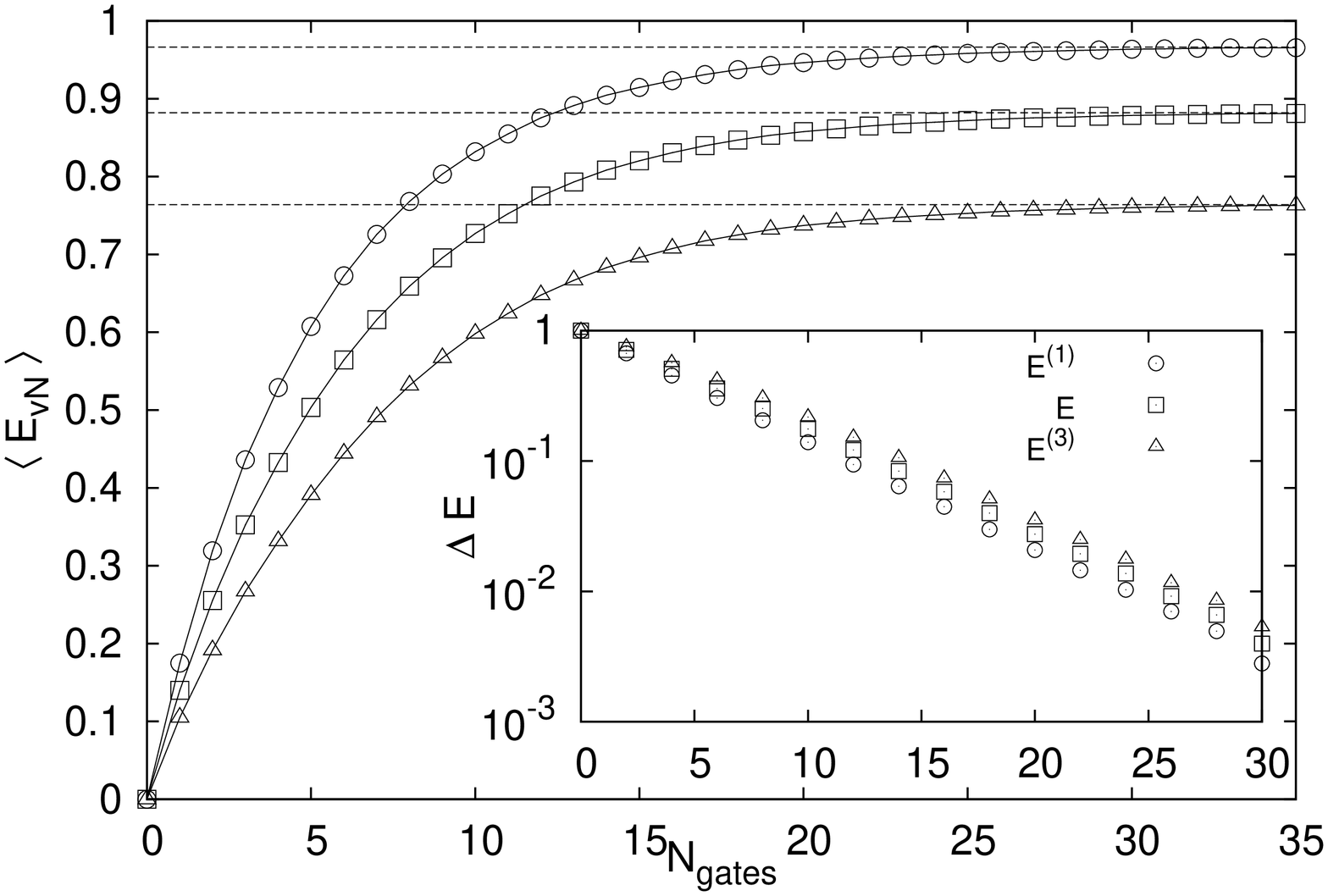}
\caption{Typical numerical simulation using the random circuit in
a system of 6-qubit. The entanglement average of the Haar measure,
represented by dashed lines is reached in $N_{gates}$ steps. Top
linear entropy as a measure of entanglement, bottom von Neumann
entropy. We plot the global entanglement (squares) and the
entanglement for the most balanced (triangles) and most unbalanced
(circles) bipartitions of the system. Inset: Decay rates for the
different entanglement measures, the rate is the same for the
linear entropy and little differences are observed for the von
Neumann entropy for different bipartitions of the system.}
\end{center}
\label{Fig1}
\end{figure}

We use two types of entanglement measures, $E_L$ and $E_{vN}$,
respectively based on two different measures for the mixedness of
the marginal density matrices $\rho_i$ associated with the
bi-partitions:
\begin{enumerate}
\item[(i)] the linear entropy $S_L =
\frac{2^m}{2^m-1}(1-Tr[\rho_i^2])$ and,
\item[(ii)] the von Neumann entropy $S_{vN}= - \frac{1}{m} Tr [\rho_i log
\rho_i]$.
\end{enumerate}
If one uses the linear entropy $S_L$, $E^{(1)}_L$ turns out to be
the well known Meyer-Wallach multipartite entanglement measure
\cite{MW02} that Brennen  showed to coincide with the average of
all the single-qubit linear entropies \cite{Brennen03}. This
measure was later generalized by Scott to the case in which all
possible bi-partitions of the system where considered
\cite{Scott04}.

 We study the convergence according to different
measures of multiqubit entanglement based upon bi-partitions. We
characterized the global entanglement with the average of the
bipartite entanglement measures associated with the $2^N-1$
bipartitions of the N-qubit system.

We compare the evolution towards the convergence of the global
entanglement with the evolution of the entanglement of the most
balanced bipartition $E^{([N/2])}$ and the entanglement of the
most unbalanced one $E^{(1)}$. We introduce the auxiliary
normalized quantity
\begin{equation}
\Delta E=\frac{E_{Haar}-\langle E\rangle}{E_{Haar}},
\end{equation}
which decays exponentially with the number of iterations. $\Delta
E$ will make easy the comparison between different bipartitions.
The saturation value $E_{Haar}$, is the mean value of the
entanglement of the Haar distribution given in \cite{HLW06,
Lubkin78} and $\langle E\rangle$ is the averaged entanglement over
system realizations. We choose the initial state to be the
separable state $|00\cdots 0\rangle$ without loosing generality.
The optimal time is defined as the number of gates required to
reach $\Delta E=0.01$. We fixed the two-qubit gate
$\vec{\lambda}=(\pi/4,0,0)$ which is known to be an optimal gate
in the non-local couplings case \cite{Z07}. The results are
qualitatively the same for any other choice of the gate.

Fig. 1 shows that the convergence rates do not depend on the
dimension of the chosen bipartition if the entanglement is
quantified with the linear entropy. However, there exist small
differences between the convergence rates if we use the von
Neumann entropy as the measure of entanglement. We have a little
faster convergence for the most unbalanced bipartition. This
behavior is more visible for higher dimensional states, the larger
the number of qubits, the larger the difference between the
convergence rates. These results imply that when working with the
linear entropy, it is enough to consider the convergence of the
Meyer-Wallach entanglement measure. If the von Neumann
entanglement measure is used, one should consider the convergence
of the entanglement of the most balanced bipartitions
$E^{([N/2])}$, because it is the one with a longer convergence
time.

\section{Gate complexity vs. time complexity}

In this section we are interested in the relation between the
number of gates in the circuit in order to reach the Haar
distribution and the physical time to perform this operation. As
an example we choose the Heisenberg Hamiltonian:
\begin{equation}
H=-\sum_j J_j\sigma^1_j\sigma^2_j+\sum_{a}B^a\sigma^a_z,
\label{Hamiltonian}
\end{equation}
where $J_j$ are anisotropic couplings ($j=x,y,z$), $B^a(t)$
($a=1,2$) is an external magnetic field in $z$ direction, and
$\sigma^1_j=\sigma_j\otimes I, \sigma^2_j=I\otimes\sigma_j$, with
$\sigma_j$, as before, the Pauli matrices. The local magnetic terms appearing in (\ref{Hamiltonian}) are needed to perform the two-qubit gate operation and are not related to the random single qubit rotations $V_{i,j}$ introduced in Sec. II.

The optimal unitary evolution operator for the Heisenberg
Hamiltonian (\ref{Hamiltonian}) was obtained in \cite{CHKO07}.
Using the results of the quantum brachistochrone formalism the
optimal time for some particular gates was calculated.

We focus our efforts in the optimal entangler gate
 $U_{\phi}$:
\begin{equation}
U_{\phi}:=
\begin{pmatrix}
  \cos \phi & 0 & 0 & \sin \phi \\
  0 & 1 & 0 & 0 \\ \label{entangler-gate}
  0 & 0 & 1 & 0 \\
  -\sin \phi & 0 & 0 & \cos \phi
\end{pmatrix} ,
\end{equation}
with the angle $\phi\in [0,\pi]$. This gate applied to an initial
separable state produces a $\phi$ dependent entangled state. When
this gate is applied to the separable state $|00\rangle$, produces
a maximally entangled state. The optimal time duration to
implement the entangler gate assuming the finite energy condition
is given by $\omega t_{\phi}=\pi\sqrt{x(1-x/2)}$, where
$x=\phi/\pi$ and $\omega$ is a constant given by the constraint.
Since the only effect of the external parameter $\omega$ is to
rescale the time, in the following we take $\omega=1$ as this
choice will not modify our results. We study the convergence rate
to the Haar distribution of this gate in terms of the physical
time. This physical time is obtained as the product of the total
number of gates $N_{gates}$ times the optimal time $t_{\phi}$
($t_{phys}=N_{gates}\times t_{\phi}$). As in this case the final
result is not affected by the choice of the measure of
entanglement, we determine the convergence using $E_L$ as a
measure for the mixedness of the marginal density matrices. As
before we consider the converge is reached for $\Delta E= 0.01$

\begin{figure}
\begin{center}
\vspace{0.5cm}
\includegraphics[scale=0.35,angle=-90]{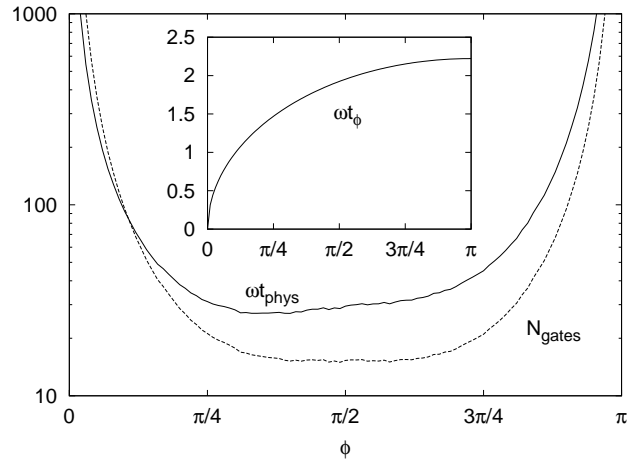}
\caption{ Number of gates (dashed line) and physical time (solid
line) to reach the convergence to the  mean value of the Haar
entanglement as a function of the gate parameter $\phi$ for a
system of 4 qubits. Inset: Optimal time to implement the entangler
$U_{\phi}$ gate as a function of $\phi$. As we have chosen units in which Planck's constant $\hbar$ is equal to one, all depicted quantities are dimensionless.}
\end{center}
\label{Fig2}
\end{figure}

Fig. 2 shows that a gate as simple as $U_{\phi}$, which depends
just on the parameter $\phi$, is enough to reveal the differences
between the gate-complexity and time-complexity concepts. The
curve $N_{gates}$ can be viewed as the time that would be
necessary to reach the convergence if all the gates took the same
time $t_{\phi} = 1$ to perform each evolution. In such case, all
gates with values of $\phi$ in the interval ($\pi / 4,3 \pi /4$)
would contribute to a reasonably efficient algorithm. The optimal
gate would be attained by $\phi = \pi / 2$. But, as can be seen in
the inset of Fig. 2, the optimal time $t_{\phi}$ needed to perform
each iteration is far from being the same for all the values of
$\phi$. It increases with the value of $\phi$, been negligible for
values of $\phi$ near zero, point at which the entangling gate
becomes the identity. If we combine both magnitudes, we obtain the
total physical time $t_{phys}$, which inherits the main properties
of $N_{gates}$ and $t_{\phi}$. The global behavior resembles that
of $N_{gates}$, both of them diverge for the extremal values $\phi
= 0$ and $\phi = \pi$ while attaining their lower and optimum
values for central values of $\phi$. The asymmetry of $t_{phys}$
compared to $N_{gates}$ comes from the behavior of $t_{\phi}$.
While $N_{gates}$ is symmetric respect its optimum and central
value $\phi = \pi / 2$, the optimum gate according to  $t_{phys}$
corresponds to a lower value of such parameter ($\phi \approx
\pi/3$).

The case studied in this section is a good example of the role
that the time complexity can play when designing an efficient
quantum algorithm According to the gate complexity there exist a
big range of values of $\phi$ for which the random states
generating algorithm is quite efficient. If one introduces the
time complexity argument this degeneracy is broken. The difference
between the time needed by these efficient gates is not huge but
it is enough to be taken into account, specially in a situation
where the algorithm must be run a large number of times.

\section{Summary and discussion}

We have studied the generation of multipartite entangled states
considering a protocol of a two-qubit fixed gate combined with two
one-qubit random rotations. A comparison was made between the gate
complexity and the time complexity revealing that they are two
different optimality problems. In a real case both complexities
should be taken into account, and the optimal gate would be one
with a reasonable good behavior according to both points of view.
Nevertheless, when the algorithm must be run a large number of
time, the optimization of time complexity is mandatory. In the
example studied in Sec. III, the time complexity optimization
allowed us to find an optimal gate between the whole family of
gates which are almost equally efficient according to the gate
complexity.

The quantum brachistochrone formalism seems to be a promising
approach for the treatment of the time complexity problem.
Following the work of Carlini {\it et al}. \cite{CHKO07} it is
possible to obtain the optimal way to realize a given quantum
gate, introducing in this derivation any constraint given by the
experimental setup.

We also focused in the possible dependence of the resulting
entanglement on the different allowed bipartitions of the system.
We found that when the linear entropy is considered, the
convergence rates are independent of the bipartition scheme.
However the convergence rates are different if we use the von
Neumann entropy. In this case the most balanced bipartition should
be considered to guarantee the convergence. These results are independent of the possible, local or non local, geometries of the system.

\begin{acknowledgements}

This work was partially supported by the MEC grant FIS2005-02796
(Spain) and by FEDER (EU). AB  acknowledges support from MEC through FPU fellowship
AP-2004-2962 and APM acknowledges support of MEC contract
SB-2006-0165.
\end{acknowledgements}

\end{document}